\documentclass{jetpl}
\twocolumn
\lat

\usepackage{amsmath}
\usepackage{amssymb}
\usepackage{amsfonts}
\usepackage{graphicx}
\usepackage{color}
\usepackage{bm}
\usepackage{cite}

\DeclareMathOperator{\Tr}{Tr} 
\DeclareMathOperator{\tr}{tr}
\DeclareMathOperator{\sgn}{sgn} 

\newcommand{\nup}{n_{\uparrow}}
\newcommand{\ndn}{n_{\downarrow}}
\newcommand{\ve}{\varepsilon}

\begin{document}

\title{Spin and Charge Correlations in Quantum Dots: An Exact Solution }

\rtitle{Spin and Charge Correlations in Quantum Dots: An Exact Solution}
\sodtitle{Spin and Charge Correlations in Quantum Dots: An Exact Solution}

\author{I.S. Burmistrov$^{+,\ddag}$, Yuval Gefen$^{\Box}$, and M.N. Kiselev$^{\triangle}$
\thanks{e-mail: burmi@itp.ac.ru, Yuval.Gefen@weizmann.ac.il, mkiselev@ictp.it}}
\rauthor{I.S. Burmistrov, Yuval Gefen, and M.N. Kiselev}
\sodauthor{I.S. Burmistrov, Yuval Gefen, and M.N. Kiselev}

\dates{\today}{*}

\address{
$^{+}$ L.D. Landau Institute for Theoretical Physics RAS,
Kosygina street 2, 119334 Moscow, Russia\\
$^{\ddag}$ Department of Theoretical Physics, Moscow Institute of
Physics and Technology, 141700 Moscow, Russia\\
$^{\Box}$ Department of Condensed Matter Physics, The Weizmann Institute of Science, Rehovot 76100, Israel \\
$^\triangle$  International Center for Theoretical Physics, Strada Costiera 11, 34014 Trieste, Italy}

\abstract{
The inclusion of charging and spin-exchange interactions within the
Universal Hamiltonian description of quantum dots  is challenging as
it leads to a non-Abelian action.  Here we present  an {\it exact}
analytical solution of the probem, in particular, in the vicinity of
the Stoner instabilty point. We   calculate several observables,
including  the tunneling density of states (TDOS) and the spin
susceptibility. Near the instability point the TDOS  exhibits  a
non-monotonous behavior as function of the tunneling energy, even at
temperatures higher than the exchange energy. Our approach is
generalizable to a broad set of observables, including the a.c.
susceptibility and the absorption spectrum for anisotropic spin
interaction.  Our results could be tested in nearly ferromagnetic
materials.
}



\maketitle

The physics  of  quantum dots (QDs)  is  a focal point of research in nanoelectronics. The introduction of the
 ``Universal Hamiltonian''~\cite{ABG,KAA}  made it possible
to simplify  in a controlled way
the intricate electron-electron interactions within  a QD.
This provided one with a  convenient framework to calculate physical observables. Within this scheme interactions
 are represented as the sum of three spatially independent terms: charging, spin-exchange, and Cooper channel.
 Notably, even the inclusion of the first two terms turned out to be non-trivial: the resulting action is
 non-Abelian ~\cite{YangMills,KG}. Attempts to account for those interactions in transport involved a
rate equation analysis~\cite{RuppAlhassid,UsajBaranger}
 and  a perturbation expansion~\cite{KG}.
Alhassid and Rupp \cite{RuppAlhassid} have analyzed some aspects of the problem (see below) exactly.
 It is known that in the presence of significant spin-exchange interaction such systems can become Stoner
 unstable. More precisely,  one distinguishes 3 regimes of behavior as
 function of increasing the strength of the exchange interaction:
 paramagnetic (no zero field magnetization), mesoscopic
 Stoner regime (finite magnetization whose value increases stepwise with the exchange)
 and  thermodynamic ferromagnetic phase
 (magnetization is proportional to the volume)~\cite{KAA}. Both the mesoscopic
 and thermodynamic  phases  manifest (Stoner)
 instabilities  towards ferromagnetic ordering. The presence of
 enhanced quantum and statistical fluctuations underlying such
 instabilities  calls for
a full-fledged quantum mechanical treatment of the problem.

Here we present  an {\it exact}   analytic algorithm  to tackle this
challenging problem. We employ our approach to a few physical
variables within the mesoscopic Stoner regime, but it can be used to
tackle the broad range of problems involving spin and charge on a
QD, and be extended to the thermodynamic ferromagnetic regime too.
As examples we  calculate the following quantities: the partition
function, the magnetic susceptibility, the distribution function of
the total spin, the tunneling density of states (TDOS), and  the
sequential tunneling conductance.  Our approach allows us to obtain
analytic results as one approaches the Stoner instability. Below we
list possible applications of our method to other physical
observables and extensions beyond the Universal Hamiltonian. The
physics discussed here can
be best tested in quantum dots with materials which are close to the
thermodynamic Stoner instability, e.g., Co impurities in Pd or Pt
host, Fe dissolved in various transition metal alloys, Ni impurities
in Pd host, and Co in Fe grains, as well as new nearly ferromagnetic
rare earth materials ~\cite{Exp:Co_in_Pt,Canfield}.

The  main reason why, in this context of a QD, the  treatment of the
exchange term is non-trivial, is the non-Abelian nature of the
action. One needs to tackle time ordered integrals of the form
\begin{equation}
\label{eq_A}
 \mathcal{A}_\gamma^{(p)} = \mathcal{T} \exp \left ( i \int_0^{t_p}
dt^\prime\, \bm{\theta}_p \bm{s}_\gamma\right ) .
\end{equation}
Here $\bm{\theta}_p$ is a dynamical, quantum  field operating on the
spin $\bm{s}_\gamma$ (whose $x$ component is  proportional to the
Pauli matrix $\sigma_x$ etc.); $p$ and $\gamma$ are indices to be
elaborated below; $\mathcal{T}$ is a time ordering operation. Wei
and Norman~\cite{WeiNorman}, addressing the problem of a quantum
spin subject to a prescribed classical time-dependent magnetic
field, have elegantly shown that by preforming a {\it non-linear}
transformation from $\theta_p^x, \theta_p^y, \theta_p^z$ to a set of
other variables (cf. Eq.~\eqref{VarChg}), Eq.~\eqref{eq_A} can be
written as a product of 3 Abelian terms (cf. Eq.~\eqref{A2P}). Even
so, that problem could not be solved. The problem of a quantum field
appears to be even more intricate.  To solve it  we employ here  a
generalized Wei-Norman-Kolokolov (WNK) method ~\cite{Kolokolov}.

We consider a quantum dot of linear size $L$ in the so-called metallic regime,
whose dimensionless conductance  $g_{\rm Th} = E_{\rm Th}/\delta \gg 1$. Here  $E_{\rm Th}$ is the Thouless energy and    $\delta$ is the (spinless) mean single particle level spacing. We account for  the following terms of the Universal Hamiltonian
\begin{equation}
H =H_0 + H_C+H_S,\qquad H_0= \sum\limits_{\alpha,\sigma} \epsilon_\alpha a^\dag_{\alpha,\sigma} a_{\alpha,\sigma} . \label{EqUnivHam}
\end{equation}
Here, $\epsilon_\alpha$ denotes the spin ($\sigma$) degenerate single particle  levels. The charging interaction $H_C=E_c \left ( \hat{n}  -N_0\right )^2$ accounts for the Coulomb blockade, with  $\hat n \equiv \sum_\alpha \hat n_\alpha=\sum_{\alpha,\sigma} a^\dag_{\alpha,\sigma} a_{\alpha,\sigma}$  being the particle number
operator; $N_0$ represents the positive background charge. The term $H_S = -J \bm{S}^2$ represents spin interactions
within the dot ($\bm{S}=\sum_{\alpha} \bm{s}_\alpha=\frac{1}{2}\sum_{\alpha}  a^\dag_{\alpha,\sigma} \bm{\sigma}_{\sigma\sigma^\prime} a_{\alpha,\sigma^\prime}$), with the components of $\bm{\sigma}$ comprising of the Pauli matrices.

The imaginary time action for this system reads:
\begin{equation}
S_{\rm tot} = \int_0^\beta \mathcal{L}  d\tau  = \int_0^\beta \Bigl [ \sum\limits_{\alpha}
\bar{\Psi}_{\alpha} (\partial_\tau + \mu)\Psi_{\alpha} - H \Bigr ] d\tau .
\notag
\end{equation}
Here $\mu$ is the chemical potential, $\beta=1/T$, $T$ the temperature, and we have introduced the Grassmann variables $\bar\Psi_{\alpha} = (\bar\psi_{\alpha\uparrow},\bar\psi_{\alpha\downarrow})^T, \Psi_{\alpha} = (\psi_{\alpha\uparrow},\psi_{\alpha\downarrow})$ to represent electrons on the dot.

Employing a Hubbard-Stratonovich transformation leads to a bosonized form
\begin{eqnarray}
\mathcal{L} &=& \sum_{\alpha}
\bar\Psi_{\alpha} \left [ \partial_\tau - \epsilon_\alpha +\mu  +i \phi+\frac{\bm{\sigma}\cdot \bm{\Phi}}{2} \right ] \Psi_{\alpha}  \notag
\\
&& + \frac{\bm{\Phi}^2}{4J} +\frac{\phi^2}{4E_c}-i N_0 \phi \notag
\end{eqnarray}
where $\phi$ and $\bm{\Phi}$ are scalar and vector bosonic fields
respectively. The  $SU(2)$ non-Abelian character of the action poses
a serious difficulty. It
 prevents one from performing a gauge transformation  \cite{KamenevGefen}
 which works efficiently in the Abelian $U(1)$ (charging only) case~\cite{KamenevGefen,EfetovTscherisch,SeldmayrLY}.
 Employing the Wei-Norman-Kolokolov trick we are able to overcome
 this difficulty.

\emph{Results. ---} Below we present our main results. The TDOS is given
by the following \emph{exact} expression
\begin{gather}
\nu(\ve) = \frac{1+e^{-\beta\ve}}{Z} \sum_{n_{\uparrow,\downarrow}\in \mathbb{Z}} e^{-\beta E_c(n-N_0)^2+\beta\mu n +\beta J m(m+1)}  \notag \hspace{.8cm}\,{}\\
\hspace{0.5cm}\times  \sum_{\alpha} \delta\Bigl [\ve-\epsilon_\alpha+\mu -E_c(2n-2N_0+1)-J \bigl (m+\frac{1}{4}\bigr )\Bigr ]\notag \\
\times
\Biggl \{2m \Bigl [ Z_{\nup}(\epsilon_\alpha)Z_{\ndn}-Z_{\nup+1}Z_{\ndn-1}(\epsilon_\alpha)\Bigr ]\hspace{0.7cm}\,{}
\notag \\ + (2m+1) \Bigl [
Z_{\nup}Z_{\ndn}(\epsilon_\alpha)-Z_{\nup}(\epsilon_\alpha)Z_{\ndn}\Bigr ]\Biggr \} .\hspace{0.35cm}\, \label{TDOS_Gen}
\end{gather}
Here $\nup(\ndn)$ represents the number of spin-up (spin-down)
electrons, the total number of electrons  $n=\nup+\ndn$,
$m=(\nup-\ndn)/2$. Note that for $m \geqslant  0$ ($m<0$) the total
spin  $S=m$ ($S=-m-1$) respectively. The normalization factor
\begin{equation}
Z=\!\!\!\!\! \sum_{n_{\uparrow,\downarrow}\in \mathbb{Z}}\!\!\!(2m
+1) Z_{\nup} Z_{\ndn}   e^{-\beta[ E_c(n-N_0)^2-\mu n- Jm(m+1)]}
\label{GCPF_Gen}
\end{equation}
coincides with the grand canonical partition function for the
Hamiltonian~\eqref{EqUnivHam}~\cite{RuppAlhassid}. The quantity
$Z_N \equiv\int_0^{2\pi} \frac{d\theta}{2\pi} e^{-i\theta N}
\prod_{\gamma} \left( 1+e^{i\theta-\beta\epsilon_\gamma}\right) $
is the canonical partition function of $N$ noninteracting spinless electrons, and
$Z_N(\epsilon_\alpha) \equiv \int_0^{2\pi} \frac{d\theta}{2\pi} e^{-i\theta N} \prod_{\gamma\neq \alpha} \left( 1+e^{i\theta-\beta\epsilon_\gamma}\right)$
determines the canonical partition function of a system of $N$
noninteracting spinless electrons under the constraint that level
$\alpha$ is not occupied.

Eqs.~\eqref{TDOS_Gen} and \eqref{GCPF_Gen} allow us to study a host
of physical observables for a given spectrum of single-particle
levels $\{\epsilon_\alpha\}$. At low temperatures, $T\lesssim
\delta$, these observables are sensitive to details of the spectrum;
their statistical averages would depend on the symmetry group of the
spectral distribution~\cite{Future}.

We now  discuss a few quantities of interest. The static spin
susceptibility can be computed as $\chi = (1/3)
\partial\ln Z/\partial J$. At high temperatures, $\delta \ll T \ll \mu /  \ln (J_\star/T)$, $J_\star=J\delta/(\delta-J)$, the
average static spin susceptibility  is given by
\begin{equation}
\label{barchi}
 \overline{\chi} = \frac{1}{2} \frac{1}{\delta-J} +
\frac{1}{12T} \frac{\delta^2}{(\delta-J)^2} - \frac{1}{12T} .
\end{equation}
This expression, underlining  the divergence at the Stoner
instability point, differs from that found by Kurland et al.~\cite{KAA}~\footnote{The average static spin susceptibility has been calculated in Ref.~\cite{KAA} near the Stoner instability, $\delta-J\ll\delta$. In our notations, the result of Ref.~\cite{KAA} at $T\gg J_\star$ becomes
 (see Eqs.(4.8), (4.13b), (4.15) of Ref.~\cite{KAA})
\begin{equation} 
\overline{\chi}  = \frac{c_0}{\delta-J}\left [  1 + c_1 \frac{\sqrt{J_\star}}{\sqrt{T}} + c_2 \frac{J_\star}{T}+\dots\right] \notag
\end{equation}
where numerical coefficients $c_0=1/3$, $c_1= \sqrt{\pi}/4$, and  $c_2\approx 0.238$ for unitary ensemble and
 $c_0=1/3$, $c_1= \sqrt{2\pi}/4$, and  $c_2\approx 0.227$ for orthogonal ensemble. The result  of Ref.~\cite{KAA}
 contradicts our result~\eqref{barchi} in which $c_0=1/2$, $c_1=0$ and $c_2=1/6$ are independent of the ensemble statistics of the single-particle levels. 
According to Ref.~\cite{KAA}, at $T=0$, (see Eq.(4.19) of Ref.~\cite{KAA}) $\overline{\chi} \propto [\delta/(\delta-J)]^2$.
As one can see from Eq.~\eqref{barchi}, our result for $T\ll J_\star$ smoothly interpolates into the result of Ref.~\cite{KAA} for $T=0$.
} and by Schechter~\cite{Schechter}~\footnote{Our result for $\overline{\chi}$ implies that the magnetic field
tends to zero first (before, e.g., temperature). The result found
by Schechter~\cite{Schechter} is valid in the limit of vanishing
temperature but finite magnetic field (provided an additional coarse
graining is performed). Generalization of Eq.~\eqref{barchi} to
finite magnetic field resembles the result of Schechter at magnetic
fields larger than temperature~\cite{Future}.}.
Near the Stoner instability,
$\delta-J\ll\delta$, it is  the first (second) term of
Eq.~\eqref{barchi} that dominates when $T \gg J_\star$ ($T \ll
J_\star$). For $T \gg J_\star$ the susceptibility behaves like a
paramagnetic Fermi liquid (with an upward renormalized g-factor). As
the system is driven towards the Stoner instability limit one
crosses over to the low temperature regime, $T \ll J_\star$, and a
non-Fermi liquid (Curie) behavior, sets in, $\overline\chi \sim
\langle \bm{S}^2\rangle /T$, where the average spin scales as $\sqrt
{\langle \bm{S}^2\rangle} \sim J_\star/\delta$. Note that the latter
approximates the discontinuous growth of the ground state spin of a
specific single electron spectrum (e.g. uniformly spaced), when  $J/
\delta$ is increased in the mesoscopic Stoner regime towards $1$. No
dynamical spin response  $\chi(\omega\neq 0)$ exists unless the dot
is connected to reservoirs, or anisotropic spin interaction is
considered.

The average moments of the total spin can be found from the
partition function $Z$ as $\langle [\bm{S}^2]^k\rangle = T^k Z^{-1}
\partial^k Z/\partial J^k$. It can be characterized by the
distribution function of $\bm{S}^2$,  $\mathcal{P}_{\bm{S}^2}(x)$
which can be found from Eq.~\eqref{GCPF_Gen}.
Near the Stoner instability $\delta-J\ll\delta$, and for the same
range of temperatures as in Eq.~\eqref{barchi}, the  distribution
becomes
\begin{equation}\label{PS2}
\mathcal{P}_{\bm{S}^2}(x) = 2\sqrt{\frac{\beta \delta^4}{\pi J^3_\star}} e^{-\beta J_\star/4} \sinh(\beta\delta\sqrt{x})\,  e^{-\beta \delta^2 x/J_\star} .
\end{equation}
The broad asymmetric  non-Gaussian nature of the distribution
becomes manifest in the  high temperature limit,
and  is not due to statistical fluctuations of the single particle
levels but rather due to the effect of the exchange interaction.

\begin{figure}[t]
\centerline{\includegraphics[width=75mm]{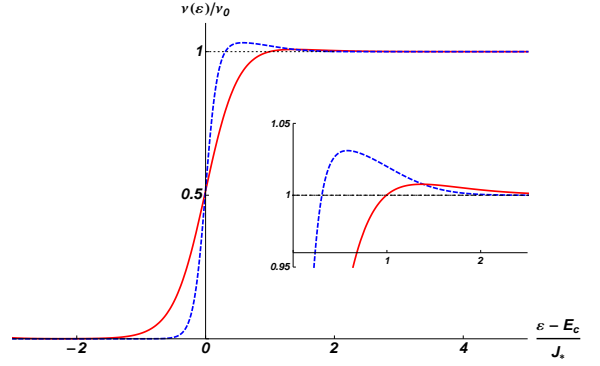}}
\caption{Fig.~\ref{FigureValley}. TDOS in the Coulomb valley. The solid
(dashed) line corresponds to $J/\delta =0.92$, $\delta/T=0.35$, and
$J_\star/T=3.95$ ($J/\delta =0.92$, $\delta/T=0.95$, and
$J_\star/T=10.70$). The inset depicts the nonmonotonic behavior.}
\label{FigureValley}
\end{figure}
%

\begin{figure}[t]
\centerline{\includegraphics[width=75mm]{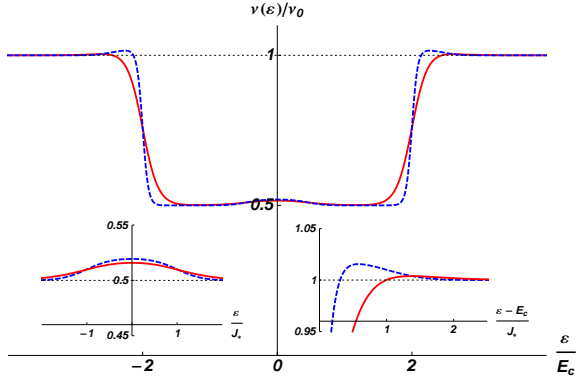}}
\caption{Fig.~\ref{FigurePeak}. TDOS at the Coulomb peak.
The parameters are the same as in Fig.\protect\ref{FigureValley}.
The insets depict the nonmonotonic behavior.} \label{FigurePeak}
\end{figure}

We next consider the average TDOS at $\delta \ll T$.   The most
interesting regime seems to be that of intermediate temperatures, $T
\ll J_\star$. Under the assumption $\mu \gg T \ln J_\star/T$,
Eq.~\eqref{TDOS_Gen} can be simplified, leading to
\begin{gather}
\frac{\overline{\nu}(\ve)}{\nu_0}=
 \sum_{n, \sigma=\pm}  e^{-\beta E_c(n-N_0)^2} \Biggl [
\left (1+\frac{J}{2J_\star}\right )
f_F(\sigma \ve - 2\sigma\Omega_n^{-\sigma}) \notag \\
- \frac{J}{2J_\star}  \mathcal{F}\left (\frac{\sigma\ve - 2\sigma\Omega_k^\sigma}{J_\star},\beta J_\star\right )\Biggr ] \Biggl /\sum_n  e^{-\beta E_c(n-N_0)^2} .
\end{gather}
Here $\Omega_n^\sigma \equiv E_c(n-N_0+\sigma/2)$, $\nu_0$ is the averaged TDOS for noninteracting electrons, and
\begin{eqnarray}
\mathcal{F}(x,y) &\equiv& \frac{1}{2}\sgn\left ( \cos\frac{\pi x}{2}\right ) e^{-\frac{y}{4}(x-1)^2 + \frac{y}{\pi^2} \cos^2 \frac{\pi x}{2}} \\
&\times& \left [ 1- \Phi\left (\frac{\sqrt{y}}{\pi} \left |\cos\frac{\pi x}{2}\right |\right)\right ]+e^{\frac{y}{2}(x-|x|)} \notag\\
&\times &\sum_{m\geqslant 0} (-1)^m e^{ - y |x| m + y m(m+1)} \theta(|x|-2m-1) .   \notag
\end{eqnarray}
$\theta(x)$ is the Heaviside step function ($\theta(0)\equiv 0$),
and the error function $\Phi(z) \equiv (2/\sqrt{\pi})\int_0^z
\exp(-t^2) dt$. As $x$ is varied for a fixed $y$, $\mathcal{F}(x,y)$
exhibits damped oscillations with a period $4$ (equivalent to an
energy scale $4J_\star$). In the limit $y\gg 1$ considered here,
these oscillations are strongly suppressed, and only the first
maximum remains visible. It leads to the appearance of a maximum in
the TDOS as illustrated in Figs.~\ref{FigureValley} and
\ref{FigurePeak}. The scaling of these oscillations with
$\sqrt{\langle S^2\rangle}\sim J_\star/\delta$ indicates that they
are due to precession of the spin of the injected electron about the
effective magnetic moment in the dot. This additional structure in
the TDOS reflects enhanced electron correlations due to the exchange
interaction. At higher temperatures, $T\gg J_\star$, there is no
interesting signature of spin-exchange on the TDOS.

One can compute the sequential conductance through the QD employing
$G=G_0 \int d\ve (-\partial f_F(\ve)/\partial
\ve)(\overline{\nu}(\ve)/\nu_0)$, where $G_0$ is the conductance of
the non-interacting QD. The maximal value of $G$ will be enhanced by
a factor $1+J/2J_{\star}$ due to the exchange term. Much more
interestingly,  the non-linear conductance at the Coulomb peak will
exhibit non-monotonic behavior, similar to Fig. ~\ref{FigurePeak}
~\cite{Future}.

\emph{Derivation. -}  Below we describe the main steps of the
derivation. Further details will be given in~\cite{Future}.

The
TDOS, $\nu(\ve) = - (1/\pi) \Imag \sum_{\alpha,\sigma}
G^R_{\alpha\sigma}(\ve)$, is determined via the imaginary part of
the retarded  Green's function, $G^R_{\alpha\sigma}(t,t^\prime) = -i
\theta(t-t^\prime)\langle
\{a_{\alpha,\sigma}(t),a^\dag_{\alpha,\sigma}(t^\prime) \}\rangle_H$
of the Hamiltonian~\eqref{EqUnivHam}. The imaginary time Green
function is given by ${G}_{\alpha\sigma}(\tau_1,\tau_2) = -
\langle\mathbb{T}_\tau \psi_{\alpha,\sigma}(\tau_1)
\bar{\psi}_{\alpha,\sigma}(\tau_2)\rangle_{S_{\rm tot}}$.

The exact one-particle Green function for the Hamiltonian~\eqref{EqUnivHam} can be written as
\begin{equation}
{G}_{\alpha\sigma}(\tau_1,\tau_2) =  \hspace{-0.1cm} \int\limits_{-\pi T}^{\pi T}\hspace{-0.1cm}d\phi_0 \,\frac{\mathcal{Z}(\phi_0)}{Z}
{D}(\tau_{12},\phi_0)  \mathcal{G}_{\alpha\sigma}(\tau_{12},\phi_0),\label{CSSep1}
\end{equation}
where $\tau_{12}=\tau_1-\tau_2$, $\phi_0$ is the static component of $\phi$, the grand canonical partition function $Z = \int_{-\pi T}^{\pi T}d\phi_0 \, {D}(0,\phi_0) \mathcal{Z}(\phi_0)$, and the so-called Coulomb-boson propagator reads~\cite{KamenevGefen,SeldmayrLY}
\begin{equation}
D(\tau,\phi_0) = e^{-E_c|\tau|} \sum_{k\in \mathbb{Z}} e^{i\phi_0 (\beta k+\tau)-\beta E_c(k-N_0)^2-2 E_c(k-N_0)\tau}.
\notag
\end{equation}
The one-particle Green function $\mathcal{G}_{\alpha\sigma}(\tau_1,\tau_2,\phi_0) $ appearing in Eq.~\eqref{CSSep1} is defined as $\mathcal{G}_{\alpha\sigma}(\tau_1,\tau_2,\phi_0) = - \langle\mathbb{T}_\tau \psi_{\alpha,\sigma}(\tau_1) \bar{\psi}_{\alpha,\sigma}(\tau_2)\rangle_\mathcal{S}$. Average is taken with respect to the action
$\mathcal{S} = \int_0^\beta d\tau \Bigl [ \sum_{\alpha} \bar{\Psi}_{\alpha} \partial_\tau \Psi_{\alpha} - \mathcal{H} \Bigr ]$.
Here $\mathcal{H} = H_0 +H_S$ with $H_0$ in which $\epsilon_\alpha$ is replaced by $\tilde{\epsilon}_\alpha=\epsilon_\alpha -\mu+i \phi_0$.
Remarkably, the charge and spin degrees of freedom are almost disentangled in the action $\mathcal{S}$. The latter involves only the spin-interaction part of the Hamiltonian~\eqref{EqUnivHam}. Traces of the charging-interaction are encoded in the variable $\phi_0$, leading to a small imaginary shift of the chemical potential.
Subsequently, the one-particle Green function can be written as
\begin{gather}
\mathcal{G}_{\alpha\sigma}(\tau_1>\tau_2) = -\mathcal{Z}^{-1}
\mathcal{K}_{\alpha\sigma}(-i\tau_{12},-i\tau_{12}+i\beta) \notag\\
\mathcal{K}_{\alpha\sigma}(t_+,t_-) = \Tr e^{-i t_+ \mathcal{H}} a^\dag_{\alpha,\sigma}  e^{i t_- \mathcal{H}}  a_{\alpha,\sigma}
\label{Eq11}
\end{gather}
and $\mathcal{Z}(\phi_0) = \Tr \exp(-\beta \mathcal{H})$. In order to evaluate the trace we perform Hubbard-Stratonovich transformations of the terms $e^{\mp i t_{\pm} H_S}$ in the evolution operators and obtain
\begin{gather}
\mathcal{K}_{\alpha\sigma}(t_+,t_-) = \prod_{p=\pm} \int \mathcal{D}
[\bm{\theta}_p]e^{- \frac{i p}{4J} \int_0^{t_p} dt^\prime\, \bm{\theta}_p^2} \label{K1}\\
\times \Tr \left [ e^{-i t_+ H_0} \prod_\gamma
\mathcal{A}_\gamma^{(+)}
a^\dag_{\alpha,\sigma}  e^{i t_- H_0}
\prod_\eta
\mathcal{A}_\eta^{(-)}  a_{\alpha,\sigma}\right ]. \notag
\end{gather}
Here $\mathcal{A}_\gamma^{(p)}$ is defined in Eq. \eqref{eq_A}.  We
have defined the bosonic fields $\bm{\theta}_p$, $p=\pm$. In order
to employ the WNK trick we use a Hamiltonian evolution of our
operators rather than a path untegral representation of
$\mathcal{G}$. Note that while $\mathcal{H}$ is time independent,
the factors $\mathcal{A}_\gamma^{(p)}$ involve time ordering
($\mathcal{T}$). This is due to the non-commutativity of the
spin-operators $\bm{s}_\gamma$.

In order to overcome the intricacy of time-ordering we use the following transformation
of variables~\cite{Footnote2} in the functional integral in Eq.~\eqref{K1}~\cite{Kolokolov},
\begin{gather}
\theta_{p}^z = \rho_{p} - 2 \kappa_{p}^p\kappa_{p}^{-p},\, \frac{\theta_{p}^x- ip \theta_{p}^y}{2}=\kappa_{p}^{-p},\notag \\
\frac{\theta_{p}^x+i p \theta_{p}^y}{2}=- i p \dot{\kappa}_{p}^p +\rho_{p} \kappa_{p}^p - (\kappa_{p}^p)^2
\kappa_{p}^{-p} , \label{VarChg}
\end{gather}
which recasts the time-ordered exponent as a product of simple Abelian ones:
\begin{gather}
\mathcal{A}_\gamma^{(p)}  =  e^{p \hat s_\gamma^{-p} \kappa_{p}^p (t_{p})} e^{i \hat s_\gamma^z\int_0^{t_{p}}
dt^\prime \rho_{p}(t^\prime)}\hspace{2.5cm}{\,}\label{A2P} \\
\times \exp \left [ i \hat s_\gamma^p \int_0^{t_{p}} dt^\prime \kappa_{p}^{-p}(t^\prime) e^{- ip \int_0^{t^\prime} d\tau \rho_{p}(\tau)} dt^\prime \right ]
.\notag
\end{gather}
Here we employ the initial condition $\kappa_{p}^p(0)
=0$~\cite{WeiNorman}, and $s_\gamma^\pm = s_\gamma^x\pm i
s_\gamma^y$. We stress that Eqs.~\eqref{VarChg} and \eqref{A2P} are
valid for a general spin operator.
In order to preserve the number of field variables (three) we impose
the following constraints on the otherwise arbitrary new complex
variables: $\rho_{p}=-\rho_{p}^*$ and $\kappa_{p}^+ =
(\kappa_{p}^-)^* $. The quantity
$\mathcal{K}_{\alpha\sigma}(t_+,t_-)$ can be then evaluated as
\begin{gather}
\mathcal{K}_{\alpha\sigma}(t_+,t_-) = \prod_{p=\pm}
\int \mathcal{D}[\rho_{p}, \kappa_{p}^{p}] e^{-\frac{i p }{4J} \int_0^{t_p}dt (\rho_p^2-4i p \dot{\kappa}_p^p \kappa_p^{-p})}
 \notag 
\\
\times e^{\frac{i p}{2} \int_0^{t_p} dt \rho_p(t)} \mathcal{C}_{\alpha\sigma}(t_+,t_-) \hspace{-0.1cm}\prod_{\gamma\neq \alpha} \hspace{-0.1cm}\mathcal{B}_\gamma(t_+,t_-) ,\label{K2}
\end{gather}
with $\mathcal{C}_{\alpha\sigma}$ and $\mathcal{B}_\gamma$ given in terms of single-particle traces:
\begin{eqnarray}
\mathcal{C}_{\alpha\sigma} &=& 
\tr \Bigl [ e^{-i \ve_\alpha \hat{n}_\alpha  t_+} \mathcal{A}_\alpha^{(+)}(t_+) a^\dag_{\alpha\sigma}e^{i \ve_\alpha  \hat{n}_\alpha  t_-} \mathcal{A}_\alpha^{(-)}(t_-) a_{\alpha\sigma} \Bigr ] , \notag \\
\mathcal{B}_\gamma &=& 
\tr \Bigl  [e^{-i \ve_\gamma \hat{n}_\gamma t_+} \mathcal{A}_\gamma^{(+)}(t_+)e^{+i \ve_\gamma \hat{n}_\gamma t_-} \mathcal{A}_\gamma^{(-)}(t_-) \Bigr ] . 
\end{eqnarray}
The expression for $\mathcal{Z}$ can be obtained from Eq.~\eqref{K2} by the
substitution of $\mathcal{B}_\alpha$ for $\mathcal{C}_{\alpha\sigma}$. 
We can now evaluate the single-particle traces in $\mathcal{B}_\gamma$ and $\mathcal{C}_{\alpha\sigma}$. The fields $\kappa_p^-$, $\kappa_p^+$ appear in $\mathcal{B}_\gamma$. It turns out that the integration over $\kappa_p^p$ first, and then $\rho_p$, can be performed exactly, yielding  $K_{\alpha\uparrow} (=K_{\alpha\downarrow})$,
\begin{gather}
\mathcal{K}_{\alpha\uparrow}= \frac{e^{-\frac{\beta J}{4}-2 i\ve_\alpha t_+}}{J\sqrt{\pi\beta J}}\hspace{-0.1cm}\int\limits_{-\infty}^\infty \hspace{-0.1cm} dh \sinh (\beta h) \prod_{\gamma\neq \alpha}  \prod_{\sigma=\pm} 
\left [ 1+ e^{\beta (\sigma h-\ve_\gamma)}\right ]\notag \\
\times 
\sum_{s=\pm} e^{i\ve_\alpha t_s+\frac{i s J t_s}{4}} e^{-\frac{(2\beta h + i s J t_s)^2}{4\beta J}}  (2\beta h +i s J t_{-s}) .\label{K4}
\end{gather}
Next, we perform the integration
over $h$ in Eq.~\eqref{K4}, substitute it into Eq.~\eqref{Eq11} and calculate the exchange-only Green function,
$\mathcal{G}_{\alpha\sigma}$. Then, integrating over $\phi_0$ in Eq.~\eqref{CSSep1} we obtain the full Green's function
${G}_{\alpha\sigma}$. Employing 
the general expression~\cite{MatveevAndreev}
\begin{equation}
\nu(\ve) = -\frac{2}{\pi} \cosh \frac{\beta \ve}{2} \sum\limits_{\alpha}\int\limits_{-\infty}^\infty dt\, e^{i\ve t}  G_{\alpha\uparrow}\left ( it+\frac{\beta}{2}\right ),
\end{equation}
we, finally, find the TDOS~\eqref{TDOS_Gen}. In a similar
way we obtain the partition function $Z$~\eqref{GCPF_Gen}.

Within WNK method one may still have some freedom in selecting
regularization of the functional integrals. It is thus useful to
check the validity of our results against some benchmarks. Our
non-trivial checks are: i) Eq.(5) for $Z$ agrees with the exact
derivation in Ref.~\cite{RuppAlhassid}. ii) The
TDOS~\eqref{TDOS_Gen} satisfies the sum rule: $\int d\ve
\,\nu(\ve)f_F(\ve) = T \partial \ln Z/\partial
\mu$~\cite{Footnote1}. iii) For $J=0$ our results for the TDOS
coincide with those of Ref.~\cite{SeldmayrLY}. iv) Our results for
$Z$ and $\nu(\ve)$ agree with a direct calculation for  single and
double level  QDs.

In summary, we have addressed here the interplay of charging and
spin-exchange interactions of electrons in a metallic quantum dot.
Even within the simple  Universal Hamiltonian framework, this
problem leads to a non-Abelian action, and necessarily requires the
evaluation of non-trivial time-ordered integrals. Our method is
applicable to the vicinity of the Stoner instability (well inside
the mesoscopic Stoner unstable regime), and could be extended to the
ferromagnetic regime. Other extensions include the study of
anisotropic spin-exchange (where the non-vanishing a.c.
susceptibility, absorption and TDOS are of particular interest),
cotunneling conductance, and an explicit inclusion of the leads.

As a demonstration of the usefulness of our exact solution we have
calculated several quantities: the partition function, the magnetic
susceptibility, the distribution function of the spin, the TDOS, and
the linear and non-linear conductance at the Coulomb peak. Some of
these quantities are amenable to experimental tests. Examples: the
broad distribution of the spin would imply significant
sample-to-sample fluctuations of the measured susceptibility; the
latter can be used to determine the distance $(1-J/\delta)$ from the
Stoner instability; the relative magnitude of the predicted
non-monotonicities in the TDOS and the conductance may exceed  $~
5-10\%$ in materials close to the Stoner instability such as Pd
($J/\delta=0.83)$ or YFe$_2$Zn$_{20}$ ($J/\delta=0.94$)
~\cite{Canfield}.

Previously, Alhassid {\it et al.}
have calculated exactly the partition function, matrix elements of
$a^\dag_{\alpha\sigma}, a_{\alpha\sigma}$~\cite{RuppAlhassid}, and
many-body eigenstates which are also eigenstates of the total spin operator~\cite{AlhassidTureci}. That approach
could be employed for the calculation of other observables. Our
independent approach is more manageable for the calculation of
higher correlators, the inclusion of exchange anisotropy, as well as
to further generalizations, as indicated above.

We acknowledge useful discussions with I. Aleiner and V.Gritsev. We
thank Y. Alhassid for explaining to us his method and the results of
his analysis. We are grateful to I. Kolokolov for providing us with
notes of his calculations and a detailed explanation. This work was
supported by RFBR Grant No. 09-02-92474-MHKC, the Council for grants
of the Russian President Grant No. MK-125.2009.2, the Dynasty
Foundation, RAS Programs ``Quantum Physics of Condensed Matter",
``Fundamentals of nanotechnology and nanomaterials", CRDF, SPP 1285
``Spintronics", Minerva Foundation, German-Israel GIF, Israel
Science Foundation, and EU project GEOMDISS.


\begin{thebibliography}{100}

\bibitem{ABG} I.\,Aleiner, P.\,Brouwer, and L.\,Glazman, Phys. Rep. \textbf{358},  309
(2002); Y. Alhassid, Rev. Mod. Phys. {\bf 72}, 895 (2000).

\bibitem{KAA} I.L. Kurland, I.L. Aleiner, B.L. Altshuler, Phys. Rev. B \textbf{62}, 14886 (2000).

\bibitem{YangMills} {\it 50 years of Yang-Mills theory}, ed. by G. 't Hooft, World Scientific, Singapore (2005).

\bibitem{KG} M.N. Kiselev and Y. Gefen, Phys. Rev. Lett. {\bf 96}, 066805 (2006).

\bibitem{RuppAlhassid}  Y. Alhassid and T. Rupp, Phys. Rev. Lett. {\bf 91}, 056801 (2003).

\bibitem{UsajBaranger} G. Usaj and H. Baranager, Phys. Rev. B {\bf 67}, 121308 (2003).

\bibitem{Exp:Co_in_Pt}
P. Gambardella et al., Science {\bf 300}, 1130 (2003);
G. Mpourmpakis, G.E. Froudakis, A.N. Andriotis, M. Menon, Phys. Rev.
B {\bf 72}, 104417 (2005).

\bibitem{Canfield} S. Jia, S. L. Bud'ko, G. D. Samolyuk, P. C. Canfield, Nat. Phys.
{\bf 3}, 334 (2007).

\bibitem{WeiNorman} J. Wei and E. Norman, J. Math. Phys. {\bf 4}, 575 (1963).

\bibitem{Kolokolov} I.V. Kolokolov, Ann. Phys. (N.Y.) {\bf 202}, 165 (1990);
M. Chertkov and I.V. Kolokolov, Phys. Rev. B {\bf 51}, 3974 (1994);
Sov. Phys. JETP {\bf 79}, 1063 (1994); for a review see I.V.
Kolokolov, Int. J. Mod. Phys. B {\bf 10}, 2189 (1996).

\bibitem{KamenevGefen} A. Kamenev and Y. Gefen, Phys. Rev. B {\bf 54}, 5428 (1996).

\bibitem{EfetovTscherisch} K.B.\,Efetov and A.\,Tschersich, Phys. Rev. B {\bf 67}, 174205
(2003).

\bibitem{SeldmayrLY} N. Sedlmayr, I.V. Yurkevich, I.V. Lerner, Europhys. Lett. \textbf{76}, 109 (2006).



\bibitem{Future} I. Burmistrov, Y. Gefen, M. Kiselev, and L. Medvedovsky, in preparation.

\bibitem{Schechter} M. Schechter, Phys. Rev. B {\bf 70}, 024521 (2004).

\bibitem{Footnote2} The Jacobian of the transformation is given as
$\mathcal{J} = \prod_{p=\pm} \exp  \frac{i p}{2} \int_0^{t_p} dt
\rho_p(t) $.

\bibitem{Footnote1} The sum rule is the consequence of the following relations, $\langle \hat n\rangle = -i G^<(t,t)$, and $G^<(\ve) = 2 i f_F(\ve) \Imag G^R(\ve)$.


\bibitem{MatveevAndreev} K.A. Matveev and A.V. Andreev, Phys. Rev. B \textbf{66}, 045301 (2002).

\bibitem{AlhassidTureci} H.E. T\"{u}reci and Y. Alhassid, Phys. Rev. B {\bf 74}, 165333 (2006).

\end{thebibliography}
\end{document}